\documentclass[12pt]{iopart}

\usepackage{graphicx}
\usepackage{dcolumn}
\usepackage{bm}
\usepackage{amssymb}
\usepackage{epsf}
\usepackage{color}
\usepackage[utf8]{inputenc}

\begin{document}

\title[Unusual magnetism of UO$_2$]{Interesting magnetic response of the nuclear fuel material UO$_2$}

\author{Sudip Pal$^1$, L. S. Sharath Chandra$^2$, M. K. Chattopadhyay$^{2,3}$ and S. B. Roy$^1$}

\address{$^1$  UGC DAE Consortium for Scientific Research, University Campus \\
Khandwa Road, Indore 452001, India}

\address{$^2$ Free Electron Laser Utilization Laboratory, Raja Rammana Centre for Advanced Technology, Indore-452 013, India }

\address{$^3$ Homi Bhabha National Institute, Training School Complex, Anushakti Nagar, Mumbai 400 094, India}
\ead{sbroy@csr.res.in}

\date{\today}

\begin{abstract}
Magnetic response of uranium dioxide (UO$_2$) has been investigated through temperature and magnetic field dependent dc magnetization measurements. UO$_2$ is a paramagnet at room temperature. The magnetic susceptibility, however, deviates from Curie-Weiss (CW) like paramagnetic behavior below $T$ = 280 K. Further down the temperature UO$_2$ undergoes phase transition to an antiferromagnetic state below $T_N$ = 30.6 K. The zero field cooled (ZFC) and field cooled (FC) magnetizations exhibit some distinct thermomagnetic irreversibility below $T_N$. The temperature dependence of the FC magnetization is more like a ferromagnet, whereas ZFC magnetization exhibits distinct structures not usually observed in the antiferromagnets. In low applied magnetic field this thermomagnetic irreversibility in magnetization exists in a subtle way even in the paramagnetic regime above $T_N$ up to a fairly high temperature, but vanishes in high applied magnetic fields. Deviation from CW law and irreversibility between ZFC and FC magnetization indicate that  the paramagnetic state above $T_N$ is not a trivial one. Magnetic response below $T_N$ changes significantly with the increase in the applied magnetic field. Thermomagnetic irreversibility in magnetization initially increases with the increase in the strength of applied magnetic field, but then gets reduced in the high applied fields. A subtle signature of a magnetic field induced phase transition is also observed in the isothermal magnetic field vartaion of magnetization. All these experimetal results highlight the non-trivial nature of the antiferromagnetic state in UO$_2$

\end{abstract}
                             

\section{Introduction}

Uranium di-oxide (UO$_2$) is a well known nuclear fuel material and used worldwide in nuclear reactors for electrical power generation and research. UO$_2$ is also recognized as a Mott-Hubbard insulator {\color{blue}\cite{Yin2008, Yu2011}}, and it promises other technological applications apart from a nuclear fuel {\color{blue}\cite{Schoenes1978,SBR2019}}. Thermal conductivity is very important for the removal of heat produced in a nuclear reactor  by fission in the nuclear fuel materials. As a result thermal properties of UO$_2$ particularly have drawn much attention over the years {\color{blue}\cite{Yin2008,Book,Harding1989,Lucuta1996,Gofryk2014}}. UO$_2$ crystallizes in face centred cubic (fcc) calcium fluorite structure ($Fm\overline{3}m$), in which U$^{4+}$ ions are surrounded by eight O$^{2-}$ ions forming a cube {\color{blue}\cite{Jaime2017}}. Therefore, the anisotropic thermal conductivity reported in this compound is rather unexpected and emphasizes on the relevance of spin-phonon coupling, which is associated with the magnetic state of the system {\color{blue}\cite{Gofryk2014}}.  The Mott insulating state in UO$_2$ further highlights the importance of strong electron-electron correlation in the system {\color{blue}\cite{Yu2011,SBR2019}}. Various techniques, including neutron scattering and nuclear magnetic resonance (NMR) have revealed a complex 3k- non-collinear antiferromagnetic (AFM) spin ordering below $T_N$ = 30.6 K. The transition is first order in nature and is accompanied by a small lattice distortion, predominantly in the oxygen cage {\color{blue}\cite{Burlet1986, Ogden1967,Ikushima2001}}. In cubic crystal field, the nine fold degenerate ($5f^{2}$, $J = 4$) state splits up with a 3-fold degenerate ground state, resulting into Jahn-Teller (JT) instability {\color{blue}\cite{Rahman}}.

Spin-orbit coupling, Coulomb interaction, antiferromagnetic exchange interaction and JT distortion are of comparable strength in UO$_2$. Below $T_N$, a quadrupoler ordering is established together with AFM spin ordering, facilitated by the interaction between cooperative JT distortion, antiferromagnetic exchange interaction and $5f$ quadrupoles {\color{blue}\cite{Allen,Willkins2006,Caciuffo2010,Pourovskii2019}}. While, there is some understanding of the physical properties of UO$_2$ in the antiferromagnetic state below $T_N$, characteristics of the high temperature state are not quite clear.  The nature of the high temperature state, although considered to be paramagnetic, is not so straightforward. Temperature dependence of thermal conductivity exhibits a minimum at $T_N$ and a maximum around $T$ = 220 K, which clearly highlight unusual physical state above $T_N$. Moreover, a positive magnetostriction has been observed above $T_N$, whereas magnetostriction is negative in the AFM state. Neutron scattering has provided evidence for dynamic JT distortion of the oxygen sublattice above $T_N$ {\color{blue}\cite{Faber1975, Faber1976, Caciuffo1999, Caciuffo2011}}. Temperature dependence of elastic constant above $T_N$ also exhibit unconventional behavior {\color{blue}\cite{Ogden1967}}. UO$_2$ has been reported to be showing some other interesting physical properties, such as piezomagnetism and magnetoelastic memory driven by spin-lattice interaction {\color{blue}\cite{Jaime2017}}.

Here we present a detailed study of the temperature ($T$) and magnetic field ($H$) dependence of magnetization ($M$) in UO$_2$. The results of our study highlight various hitherto unknown interesting aspects of the magnetic respose of UO$_2$.  We show that below $T$ = 280 K the low field magnetization or susceptibility deviates from standard Curie-Weiss paramagnetic behavior. Abrupt changes in both the zero field cooled (ZFC) and field cooled (FC) magnetization are observed at $T_N$ = 30.6 K where UO$_2$ undergoes a phase transition to antiferromagnetic state along with distinct thermomagnetic irreversibility i.e. $M_{ZFC} (T)$ $\neq$ $M_{FC} (T)$  inside the antiferromagnetic state. The magnetic response in the antiferromagnetic state changes considerably with the increase in the applied magnetic field. The observed behaviour may be correlated to a magnetic field induced phase transition. The thermomagnetic irreversibility in the antiferromagnetic state continues to exist in the temperature range above $T_N$ and below $T$ = 150 K, but it gets suppressed completely at higher applied magnetic fields. 

\begin{figure}[t]
\centering
\includegraphics[scale=0.28]{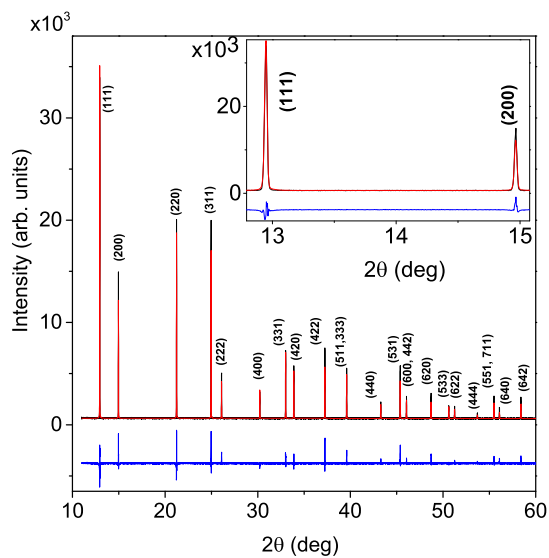}
\caption{\label{fig:epsart} X-ray diffraction pattern of UO$_2$ powder sample recorded at room temperature using wavelength, $\lambda$ = 0.7121 \AA. The data have been fitted using rietveld refinement method.  Indexes of the planes corresponding to peaks are also mentioned. The blue line at the bottom shows the difference between experimental data and fitted curve.}
\end{figure}
\section{Experimental details}
Powder sample of UO$_2$ has been prepared by reducing UO$_3$ in hydrogen atmosphere at 700$^{\circ}$C at Bhabha Atomic Research Centre, Trombay.  Room temperature X-ray diffraction data have been recorded at the wavelength of $\lambda$ = 0.7121 \AA ~ in the beamline-BL12 in Indus-2 synchrotron radiation source at Raja Ramanna Centre for Advanced technology (RRCAT), Indore, India. Magnetic measurements have been carried out in a MPMS-3 SQUID-VSM magnetometer (M/S Quantum Design, USA). Temperature dependence measurements of magnetization have been performed in temperature sweep mode of measurement at 0.5 K/min cooling and heating rate.

\section{Results and discussion}

Fig.{\color{blue}1} shows the X-ray diffraction data of UO$_2$ measured at  room temperature in $\theta-2\theta$ geometry. In the inset, we have shown a magnified view of the data around (111) peak, which is the most intense peak in this case. Sharp diffraction peaks and flat background reveal the good crystalline nature of the sample. The data have been analyzed by Rietveld refinement method using Fullprof software package. All the peaks can be indexed in cubic fcc structure (space group $Fm\overline{3}m$), which rules out the presence of any secondary phase in the sample. The optimized lattice parameter obtained from fitting is, $a$ = 5.4594 \AA. Lattice parameter of UO$_2$ is sensitive to oxygen stoichiometry of the sample and follows an empirical law given by, $a = 5.4705 - 0.132x$, where $x$ is defined as UO$_{2+x}$ {\color{blue}\cite{Teske1983}}. In our case, the obtained lattice parameter corresponds to, $x = 0.08$.  

\begin{figure*}[t]
\centering
\includegraphics[scale=0.42]{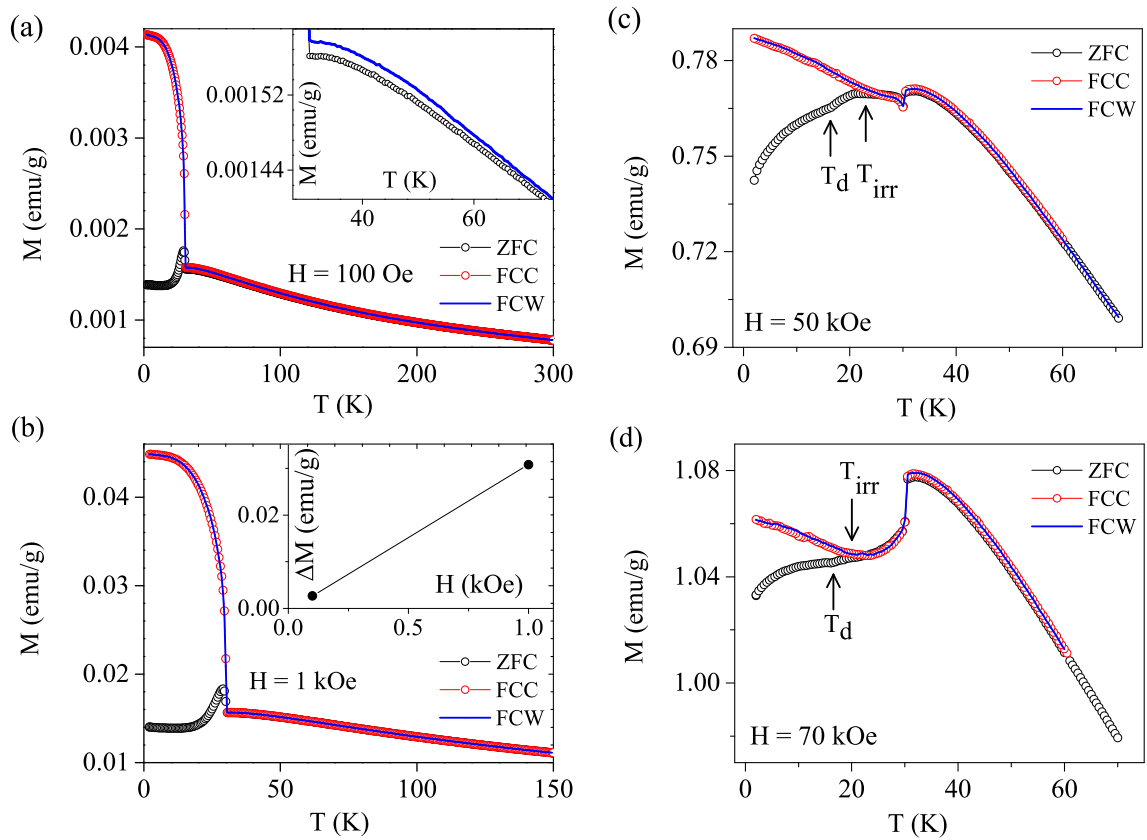}
\caption{\label{fig:epsart} (a) Magnetization ($M$) versus temperature ($T$) plots measured in applied fields of (a)  $H$ = 100 Oe, (b) $H$ = 1 kOe, (c) $H$ = 50 kOe and (d) $H$ = 70 kOe in the ZFC, FCC and FCW modes. FCC and FCW curves overlap at all temperatures. The $M_{ZFC}(T)$ and $M_{FC}(T)$ curves show bifurcation below $T_N$. The bifurcation also persists above $T_N$ at least upto 150 K in the applied fields of 100 Oe. The inset in Fig. (a) shows the magnified view of ZFC and FCW curve above $T_N$ in applied field of 100 Oe. The inset in Fig. (b) shows $\Delta M = M_{FC}-M_{ZFC}$ in fields of $H$ = 100 Oe and 1 kOe. In applied fields of 50 kOe and 70 kOe  $M_{ZFC}(T)$ and $M_{FC}(T)$ curves show a sudden and discontinuous fall in magnetization at $T_N$. $M_{ZFC}(T)$ and $M_{FC}(T)$ bifurcate below temperature $T_{irr} < T_N$ (see Figs. c and d).}
\end{figure*}

Fig. {\color{blue}2(a)} shows the temperature dependence of magnetization ($M$) in UO$_2$ measured at an applied field of $H$ = 100 Oe in the zero field cooled (ZFC) warming, field cooled cooling (FCC) and field cooled warming (FCW) modes.  In ZFC mode the measuring field $H$ = 100 Oe is applied after cooling the sample to the lowest temperature (here 2 K) of measurement in zero external field. ZFC magnetization ($M_{ZFC}$) is then measured while warming the sample. After reaching the highest temperature (room temperature) of measurement the sample is subsequently cooled back to 2 K in same field while measuring the FCC magnetization ($M_{FCC}$). After FCC measurements the FCW magnetization ($M_{FCW}$) is measured while by warming the sample again gradually to room temperature in the same field. Temperature variation of magnetization in the high temperature regime in all the modes (ZFC, FCC and FCW) indicate the presence of magnetic moment in UO$_2$, which is in line with the Mott insulator status of the sample. Below $T$ = 50 K, $M$ tends to flatten while the temperature is further reduced and then $M$ changes abruptly around $T_N$ = 30.6 K. This temperature matches well with the reported temperature where UO$_2$ supposedly undergoes a phase transition to a complex antferromagnetic state {\color{blue}\cite{Ikushima2001}}. Figure 2(a) highlights the presence of a very distinct thermomagnetic irreversibility i.e. $M_{ZFC}(T)$ $\neq$ $M_{FC}(T)$  in the antiferromagnetic state. Such thermomagnetic irreversibility is not expected in any standard antiferromagnet ${\color{blue}\cite{blundell}}$, and to the best of our knowledge has not been reported earlier for UO$_2$. M$_{FCC}(T)$ and $M_{FCW}(T)$ curves overlap at all temperatures of measurement and now onwords we will designate the field cooled magnetization as $M_{FC}$. Interestingly in an expanded scale (see the inset of Fig. {\color{blue}2(a)}) one can see that the  $M_{ZFC}(T)$ and $M_{FC}(T)$ curves actually start to bifurcate below $T$ = 150 K, which is well over $T_N$ and nearly equal to $5 T_N$; the bifurcation gradually increases with decrease in temperature.  The bifurcation between the $M_{ZFC}(T)$ and $M_{FC}(T)$ curves below the antiferromagnetic transition temperature $T_N$ is much larger than the bifurcation above $T_N$. Just below the transition temperature, $M_{ZFC}(T)$ curve starts to rise sharply with increase in temperature, which is followed by a rapid fall resulting into a narrow peak around $T$ = 29 K. Below about $T$ = 10 K, $M_{ZFC}$(T) curve increases slowly with decreasing temperature. All these features are not typical of a collinear antiferromagnet. Field cooled magnetization $M_{FC}(T)$ on the other hand rises sharply with decrease in temperature  below $T_N$ and tends to saturate below $T$ = 15 K. Such a sharp rise in magnetization followed by a tendency towards saturation is not expected in the case of a collinear AFM state and is usually a characteristic feature of a ferromagnetic state. This kind of behaviour possibly arises due to non-collinear spin ordering within the AFM state of UO$_2$, which results into weak ferromagnetic response below $T_N$.  It may be noted here that, although the antiferromagnetic  transition in UO$_2$ is reported to be first order in nature, $M_{FCC}(T)$ and $M_{FCW}(T)$ do not show any thermal hysteresis acrosss $T_N$. It means that the supercooling and superheating phenomena, which are usually observed across a first order phase transition in many systems${\color{blue}\cite{sbroyprl,royjpcm}}$, are absent across the antiferromagnetic transition in UO$_2$. 

Fig. {\color{blue}2(b)} presents the results of the magnetization measurements as a function of temperature carried out in an applied magnetic field of $H $ = 1 kOe. $M_{ZFC}$, $M_{FCC}$ and $M_{FCW}$ curves appear to be qualitatively similar to those obtained at $H$ = 100 Oe. However, there are important differences, which are noted below:  

\begin{enumerate}
\item The difference between $M_{ZFC}$ and $M_{FC}$ curves i.e. the thermomagnetic irreversibility above $T_N$ is completely erased.
\item The thermomagnetic irreversibility below $T_N$ increases by a large extent than that at $H$ = 100 Oe. The value of $\Delta M = M_{FCW}-M_{ZFC}$ at $H$ = 100 and 1 kOe is shown in the inset of Fig. {\color{blue}2(b)}.
\item Bifurcation of $M_{ZFC}$ and $M_{FC}$ curves starts just below $T_N$ and this temperature does not seem to change with increase in applied magnetic field from 100 Oe to 1 kOe. 
\item The low temperature increase in $M_{ZFC}$ is more pronounced.  
\end{enumerate}

\begin{figure}[t]
\centering
\includegraphics[scale=0.35]{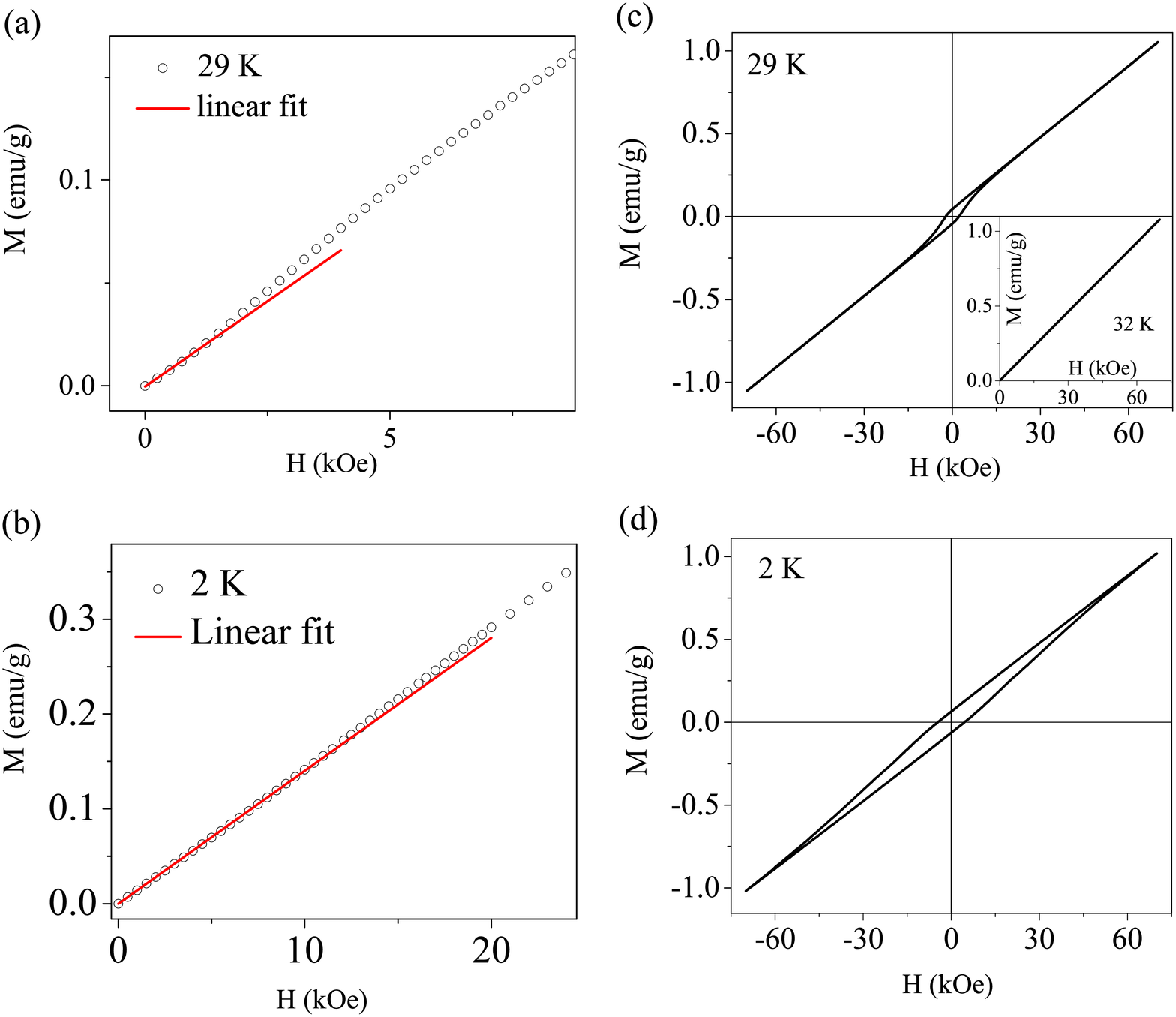}
\caption{\label{fig:epsart} Isothermal field ($H$) variation of magnetization($M$) in the antiferromagnetic regime at $T$ = 29 K and 2 K: (a) and (b) show the virgin curves at $T$ = 29 and 2 K near origin. $M(H)$ in the low field region is fitted and extrapolated to show the change in slope in the virgin $M-H$ curve.  (c) and (d) show isothermal $M-H$ loops at $T$ = 29 and 2 K. In the Inset of (c) $M-H$ curve at $T$ = 32 K has been shown only in the first quadrant.}
\end{figure}

Thermomagnetic irreversibility is a characteristic feature of spin glass or cluster glass systems, where it arises due to freezing of spins or magnetic moments originating from the competing magnetic interactions and associated frustration $\cite{Mydosh2015,Sudip2021}$. Thermomagnetic irrersivility is also observed in ferromagnets, which is associated with domain wall pinning $\cite{sbroy1}$. Even some antiferromagnets with ThCr$_2$Si$_2$ structure exhibits thermomagnetic irrersivility, which was attributed to the stacking faults in those compunds with layered structures  $\cite{sbroy2}$. However, in all these magnetic systems the thermomagnetic irreversibility gets suppressed with increase in the applied magnetic field.  This is in contrast with the observed increase in thermomagnetic irreversibility in UO$_2$ with the increase in applied fields from 100 Oe to 1 kOe. Such behaviour in UO$_2$, however, changes in the region of high applied fields. Fig.{\color{blue} 2(c) and 2(d)} present the temperature dependence of magnetization at $H$ = 50 kOe and 70 kOe measured in ZFC, FCC and FCW modes. In this case the thermomagnetic irreversibility as well as $T_{irr}$ get reduced as the applied magnetic field is increased from 50 kOe to 70 kOe, whereas the antiferromagnetic transition temperature $T_N$ remains largely unaffected. It may also be noted that the thermomagnetic irreversibility in the paramagentic regime is totally absent in this high applied field regime. 

From the results discussed above it is clear that magnetic response in low applied magnetic fields in both ZFC and FC states below $T_N$ in UO$_2$  is distinctly different from that at high applied magnetic fields. More evidences in this direction emerge if one looks carefully to the temperature dependence of high field magnetization.   As the UO$_2$ sample is cooled from high temperature, magnetization increases and tends to saturate around 35 K. However, just above $T_N$, magnetization decreases by a small amount, resulting into a small hump prior to $T_N$. The magnetic response at $T_N$, and also at low temperatures, change drastically both in the ZFC and FC states, as compared to magnetic behavior at low fields, shown in Fig. {\color{blue}2(a) and 2(b)}. At $H$ = 50 kOe, both the $M_{ZFC}(T)$ and $M_{FC}(T)$ curves undergo an abrupt and discontinuous fall at $T_N$, which is immediately followed by a small rise in magnetization with decrease in temperature. Then, they show a relatively flat region over a temperature region of 28-22 K. In the lower temperature region, the $M_{ZFC}$ and $M_{FCC}$ curves bifurcate: the $M_{ZFC}$ curve starts to decrease, whereas the $M_{FCC}$ curve slowly increases. Note that, the $M_{ZFC}$ also shows an additional shallow dip around $T_d$ = 15 K.  Note that, the bifurcation between  $M_{ZFC}(T)$ and $M_{FC}(T)$ curves appears only below a temperature $T_{irr}$ (see Fig. {\color{blue}2(c)}), which is lower than the transition temperature $T_N$. At $H$ = 70 kOe, magnetization shows larger drop at $T_N$ and all the three curves gradually decrease with further decrease in temperature  (see Fig. {\color{blue}2(d)}). Below $T_{irr}$, the $M_{ZFC}$ curve continues to decrease, whereas the $M_{FCC}/M_{FCW}$ curve increases. This is in clear contrast to the temperature dependence of magnetization obtained in applied magnetic fields of 100 Oe and 1 kOe (see Fig.2(a) and 2(b)).

To investigate further on the antiferromagnetic state in UO$_2$, we present in Fig. {\color{blue}3} the isothermal $M-H$ curves at $T$ = 29 and 2 K measured starting from the ZFC state. Here, the sample is initially cooled to the temperature of measurement in the absence of any external field. Then $M$ is measured while increasing $H$ isothermally to 70 kOe to record initial (virgin) curve, which is shown in Figs. {\color{blue}3(a)} and {\color{blue}3(b)} for $T$ = 29 and 2 K, respectively. After recording the virgin curve, $M$ is measured while varying $H$ between $\pm$70 kOe to record the envelope curves, which are shown in Figs. {\color{blue}3(c)} and {\color{blue}3(d)}. The virgin curves at both temperature lie within the envelope curves. Note that, at $T$ = 29 K, the $M-H$ curve shows a small change in the slope above $H_a$ = 1.2 kOe and continues to increase at higher field up to the highest applied field of $H$ = 70 kOe. Whereas, the virgin $M-H$ curve at $T$ = 2 K deviates from linearity above  $H_a$ = 15 kOe. The envelope $M-H$ curve at $T$ = 29 K shows a small hysteresis with a coercive field of around $H_C$ = 2.2 kOe, which increases to 4.5 kOe at 2 K. The $M-H$ curves do not show any tendency of saturation till $H$ = 70 kOe. The nonsaturating $M-H$ curve highlights the antiferromagnetic state but the presence of hysteresis is rather unexpected in AFM state. At further lower temperature, both coercive field and remanant magnetization further increase as evident from the $M-H$ curve at $T$ = 2 K.  This change in the slope of $M-H$ curves above $H = H_a$, may indicate some sort of field induced transition of the zero field cooled AFM state.

The increase in the difference between  $M_{ZFC}(T)$ and $M_{FC}(T)$ with magnetic applied magnetic field is one of the important signatures of a kinetically arrested first order transitions{\color{blue}\cite{royjpcm,Chattopadhyay2005,Banerjee2006,Roy2009,chaddahbook}}. In this case, the dynamics of a first order phase transition gets arrested in a $H-T$ window. The difference between  $M_{ZFC}(T)$ and $M_{FC}(T)$ increases because cooling in different magnetic fields produce different volume fractions of the high and low temperature magnetic phases. As stated earlier, the transition in UO$_2$ is first order in nature, and like a kinetically arrested system thermomagnetic irreversibility in the low field regime increases with the applied magnetic field. In this context, the $M-H$ curves shown in Fig. {\color{blue}3}, can be rationalized in the following manner: the ZFC state undergoes a field induced transition above a critical field $H_a$, as observed from the change in slope. However, in the field decreasing cycle, the reverse transition is not observed, so that the magnetic state induced by applied magnetic field persists during the entire envelope curve and gives rise to hysteresis loop. Such behaviour has been reported in the literature in the cases of kinetic arrest of first order phase transition in various magnetic systems, where the zero field cooled state may be either equilibrium low-$T$ phase, or the kinetically arrested high-$T$ phase, depending on the nature of the ground state {\color{blue}\cite{kranti,chaddahbook}}. Now, after zero field cooling, in the first case, when magnetic field is increased, above certain field, the equilibrium phase undergoes a field induced transition to the high-$T$ phase as the superheating band is crossed {\color{blue}\cite{Banerjee2006,chaddahbook,kranti}}. On the other hand, in the second case, the kinetically arrested state devitrify into the equilibrium low-$T$ phase while increasing the field. In either cases, the envelope curve does not show the reverse transition.

\begin{figure}[h]
\centering
\includegraphics[scale=0.36]{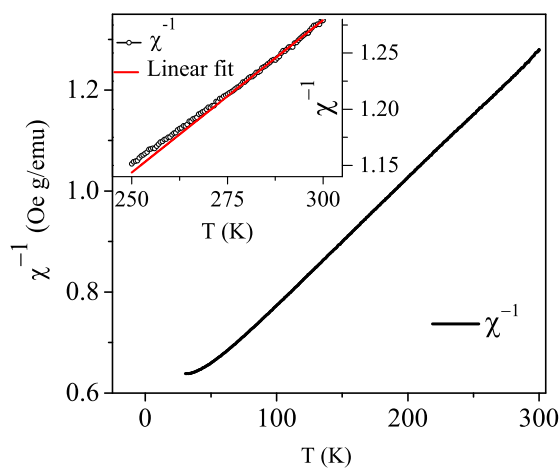}
\caption{\label{fig:epsart} $\chi^{-1}$ versus $T$ data measured at $H$ = 100 Oe in the ZFC mode. The inset shows the CW law fitting above 280 K.}
\end{figure}

The unusual dependence of magnetization on temperature and the finite difference between $M_{ZFC}(T)$ and $M_{FC}(T)$ curves much above $T_N$ clearly highlight an unconventional paramagnetic state at high temperatures. Therefore, to further understand the magnetic property of UO$_2$ at high temperatures, $\chi^{-1}$ versus $T$ at $H$ = 100 Oe measured in ZFC mode is plotted in Fig. {\color{blue}4}. The $\chi^{-1}$ versus T data appears to be grossly linear in the temperature range from $T$ = 300 K to 70 K. We tried to fit the results in this temperature regime by using Curie-Weiss (CW) law, given by $\chi = \frac {C}{T+T_0}$, where $ C = \frac {N\mu_{eff}^2}{3k_B} =\frac {Ng^2J(J+1)}{3k_B}$. However, when looked carefully, the CW law does not fit the experimental data in the entire temperature regime. In fact, it deviates from linearity below around $T$ = 280 K (see the inset of Fig. {\color{blue}4}). The effective magnetic moment $\mu_{eff}$ obtained to be 2.82 $\mu_B$/f.u., which is smaller than the expected value of 3.57 $\mu_B$/f.u. for $J$ = 4 (L =5 and S= 1). However, it matches extremely well with value of effective moment expected for the threefold degenerate ground state in UO$_2$ in a cubic crystal field {\color{blue}\cite{Ikushima2001}}. $T_0$ is obtained to be around 171 K, which is around 5.6 times higher than the transition temperature.  The deviation of the susceptibility from CW law, the presence of irreversibility much above $T_N$, higher value of $T_0$ than $T_N$ are interesting. There exist some earlier reports of the unusual characteristics above $T_N$ {\color{blue}\cite{Gofryk2014}}. In inelastic neutron scattering experiments, magnetic inelastic response has been observed above $T_N$ up to as high as $T$ = 200 K {\color{blue}\cite{Caciuffo1999}}. It has been suggested that the coherent motion of the neighbouring oxygen cages produces uncorrelated 1-k type dynamical JT distortion in the paramagnetic state of UO$_2$ {\color{blue}\cite{Caciuffo1999}}. With the lowering in temperature correlation builds up and a static 3-k type distortion condenses at $T_N$. The tendency of $M$ to saturate as $T_N$ is approached during cooling, the deviation from the CW law, existence of bifurcation between the ZFC and FC curves highlight the unusual paramagnetic state at high temperature, which may arise due to short range 1-k type  dynamic JT distortion.

\section{Conclusion}
Summarizing, we can say from the results of temperature and magnetic field dependent studies of dc magnetization  that the low temperature antiferromagnetic state in UO$_2$ is non-trivial in nature and is accompanied with large thermomagnetic irreversibility. The field cooled magnetization $M_{FC}(T)$ measured in low fields of 100 Oe and 1 kOe indicates the presence of ferromagnetic character in UO$_2$. This inference is further corroborated by the presence of hysteresis in isothermal field variation of magnetization alongwith appreciable coercive field. The nature of the thermomagnetic irreversibility changes in the presence of high applied magnetic field. In the low magnetic field region the thermomagnetic irreversibility irreversibility increases with the increase in applied field, which can be rationalized in terms of the kinetic arrest of magnetic field induced first order phase transition. The thermomagnetic irreversibility in the high magnetic field regime deceases with the increase in applied field. Prima facie this behaviour is quite similar to that observed in some ferro and antiferromagets, where it was attributed to the hindrance in domain wall motion. Furthermore the paramagetic state in UO$_2$ is also unusual with a deviation of Curie-Weiss law and the presence of thermomagnetic irreversibility in the temperature region much above $T_N$. These results indicate the existence of some short range magnetic correlations well inside the paramagnetic region of UO$_2$. Overall the results of our present study are expected to stimulate further microscopic measurements involving neutron scattering and muon spin rotation ($\mu$SR) measurents to find out the exact microscopic nature of magnetic states in various magnetic field ($H$) - temperature ($T$) regime of UO$_2$.

\section{Acknowledgment}
We acknowledge Dr. Vinay Kumar of Bhabha Atomic Research Centre, Trombay for providing us with powdered UO$_2$ sample and Dr. A. Sagdeo and Dr. A. K. Sinha of Raja Ramanna Centre for Advanced Technology, Indore for help in X-ray diffraction measurement. S B Roy acknowledges financial support from Department of Atomic Energy, India in the form of Raja Ramanna Fellowship. Sudip Pal and S B Roy  thank Dr. A J. Pal, Director, UGC-DAE CSR and Dr. D. M. Phase, Centre Director, UGC-DAE CSR, Indore Centre for suuport and encouragement.

\section{\label{sec:level}References: } 


\begin{thebibliography}{Bibliography}
\bibitem{Yin2008} Quan Yin and Sergey Y. Savrasov, Phys. Rev. Lett {\bf 100}, 225504 (2008).
\bibitem{Yu2011} S. -W. Yu et al, Phys. Rev. B {\bf 83}, 165102 (2011).
\bibitem{Schoenes1978} J. Schoenes, J. Appl. Phys. {\bf 49} 1463 (1978).
\bibitem{SBR2019} S. B. Roy, {\it{Physics of Mott insulator: Physics and applications}}, IOP Publishing (2019).
\bibitem{Book} Jack Leland Daniel, J. Matolich, H. W. Deem, {\it{ Thermal conductivity of UO$_2$}}, Hanford atomic products operation (1962).
\bibitem{Harding1989} J. H. Harding and D. G. Martin,  J. Nucl. Mater. {\bf 166}, 223 (1989).
\bibitem{Lucuta1996} P.G.Lucuta, Hj. Matzke, I.J.Hastings,  J. Nucl. Mater. {\bf 232}, 166 (1996).
\bibitem{Gofryk2014} K. Gofryk et al, Nature Communication 5:4551 (2014).
\bibitem{Jaime2017} M. Jaime et al, Nature Communication 8: 99 (2017).
\bibitem{Burlet1986} P. Burlet, J. Rossat-Mignod, S. Quezel, O. Vogt, J.C. Spirlet, and J. Rebizant, J. Less-Common Met. 121, 121 (1986).
\bibitem{Ogden1967} Ogden G. Brandt and Charles T. Walker, Phys. Rev. Lett {\bf 18}, 11 (1967).
\bibitem{Ikushima2001} K. Ikushima et al, Phys. Rev. B {\bf 63} 104404 (2001).
\bibitem{Rahman} H.U. Rahman and W.A. Runciman, J. Phys. Chem. Solids {\bf 27}, 1833 (1966); {\bf 30}, 2497 (1969); H.U. Rahman, Physica (Amsterdam) {\ }, 511 (1970).
\bibitem{Allen} S.J. Allen, Phys. Rev. 166, 530 (1968); 167, 492 (1968).
\bibitem{Willkins2006} S. B. Wilkins, R. Caciuffo, C. Detlefs, J. Rebizant, E. Colineau, F. Wastin, and G. H. Lander, Phys. Rev. B {\bf 73}, 060406(R) (2006).
\bibitem{Caciuffo2010} R. Caciuffo, G. van der Laan, L. Simonelli, T. Vitova, C. Mazzoli, M. A. Denecke, and G. H. Lander, Phys. Rev. B {\bf 81}, 195104, (2010).
\bibitem{Pourovskii2019} Leonid V. Pourovskii and Sergii Khmelevskyi, Phys. Rev. B {\bf 99} 094439 (2019).
\bibitem{Faber1975} J. Faber, G. Lander and B. Cooper, Phys. Rev. Lett. {\bf 35}, 1770 (1975).
\bibitem{Faber1976} J. Faber and G. Lander, Phys. Rev. B {\bf 14}, 1151 (1976) 
\bibitem{Caciuffo1999}  R. Caciuffo, G. Amoretti, P. Santini, G. H. Lander, J. Kulda, and P. de V. Du Plessis, Phys. Rev. B {\bf 59}, 13 892 (1999).
\bibitem{Caciuffo2011}R. Caciuffo, P. Santini, S. Carretta, G. Amoretti, A. Hiess, N. Magnani, L.-P. Regnault, and G. H. Lander, Phys. Rev. B {\bf 84}, 104409 (2011).
\bibitem{Teske1983} K. Teske, H. Ullmann, and D. Rettig, J. Nucl. Mater. {\bf 116}, 260 (1983).
\bibitem{blundell} S. J. Blundell, Magnetism in Condensed Matter (Oxford University Press, 2001).
\bibitem{sbroyprl} S. B. Roy, G. K. Perkins, M. K. Chattopadhyay, A. K. Nigam, K. J. S. Sokhey, P. Chaddah, A. D. Caplin and L. F. Cohen, Phys. Rev. Lett,. {\bf 92} 147203 (2004).
\bibitem{royjpcm} S. B, Roy, J. Phys.: Condens. Matter {\bf 25}  183201 (2013).
\bibitem{Mydosh2015} J. A. Mydosh, Rep. Prog. Phys. {\bf 78}, 052501 (2015).
\bibitem{Sudip2021} Sudip Pal, Kranti Kumar, A. Banerjee, S. B. Roy and A. K. Nigam, J. Phys.: Condens. Matter {\bf 33} (2021) 025801 (2021) and references therein.
\bibitem{sbroy1} S.B. Roy, A.K. Pradhan, P. Chaddah and B.R. Coles, Solid St. Commun. {\bf 99} 563 (1996). 
\bibitem{sbroy2} S. B. Roy, A.K. Pradhan and P. Chaddah, J. Phys.: Condens. Matter {\bf 6} 5155 (1994).
\bibitem{sbroy3}  S.B. Roy, A.K. Pradhan, P. Chaddah and E. V. Sampathkumaran, J. Phys.: Condens. Matter {\bf 9}  2465 (1997).
\bibitem{Chattopadhyay2005} M. K. Chattopadhyay, S. B. Roy, and P. Chaddah, Phys. Rev. B {\bf 72}, 180401(R) (2005).
\bibitem{Banerjee2006} A. Banerjee, K. Mukherjee, Kranti Kumar, and P. Chaddah, Phys. Rev. B {\bf 74}, 224445 (2006). 
\bibitem{Roy2009} S. B. Roy and M. K. Chattopadhyay, Phys. Rev. B  {\bf 79}, 052407 (2009).
\bibitem{chaddahbook} P. Chaddah, First Order Phase Transitions of Magnetic Materials (Taylor and Francis, 2017).
\bibitem{kranti} Kranti Kumar et al, Phys. Rev. B {\bf 73} 184435 (2006). 
\end{thebibliography}
\end{document}